\documentclass[aps,pre,nofootinbib,showpacs,tightenlines,preprint,titlepage,amsmath]{revtex4}
\usepackage{bm}

\newcommand{\cC}{\ensuremath{\mathcal{C}}}
\newcommand{\cQ}{\ensuremath{\mathcal{Q}}}
\newcommand{\cP}{\ensuremath{\mathcal{P}}}
\newcommand{\cT}{\ensuremath{\mathcal{T}}}
\newcommand{\half}{\mbox{$\textstyle{\frac{1}{2}}$}}

\begin{document}
\title{Small-$\epsilon$ behavior of the Non-Hermitian $\cP\cT$-Symmetric
Hamiltonian $H=p^2+x^2(ix)^\epsilon$}

\author{Carl M. Bender}\thanks{Permanent address: Department of Physics,
Washington University, St. Louis, MO 63130, USA}
\email{cmb@wustl.edu}
\author{Karim Besseghir}
\email{karim.besseghir07@epfl.ch}
\author{Hugh F. Jones}
\email{h.f.jones@ic.ac.uk}
\author{Xinghui Yin}
\email{xinghui.yin07@imperial.ac.uk}
\affiliation{Theoretical Physics, Blackett Laboratory, Imperial College, London
SW7 2BZ, UK}

\date{\today}

\begin{abstract} 
The energy eigenvalues of the class of non-Hermitian $\cP\cT$-symmetric
Hamiltonians $H=p^2+x^2(ix)^\epsilon$ ($\epsilon\geq0$) are real, positive, and 
discrete. The behavior of these eigenvalues has been studied perturbatively for
small $\epsilon$. However, until now no other features of $H$ have been examined
perturbatively. In this paper the small-$\epsilon$ expansion of the $\cC$
operator and the equivalent isospectral Dirac-Hermitian Hamiltonian $h$ are
derived.
\end{abstract}

\maketitle
\section{Introduction}
\label{s1}

For non-Hermitian $\cP\cT$-symmetric Hamiltonians it has been established
that the physical requirements of spectral positivity and unitarity can be met
even though the Hamiltonian is not Hermitian in the Dirac sense. (A Hamiltonian
$H$ is Hermitian in the {\it Dirac sense} if it satisfies $H=H^\dag$, where the
Dirac adjoint-symbol $\dag$ indicates combined complex conjugation and matrix
transposition.) Many $\cP\cT$-symmetric model Hamiltonians have been studied
\cite{R1}, but the first non-Hermitian $\cP\cT$-symmetric Hamiltonian for which
spectral positivity and unitarity were verified is
\begin{equation}
H=p^2+x^2(ix)^\epsilon\quad(\epsilon\geq0).
\label{e1}
\end{equation}

It was shown in 1998 that the spectrum of the class of Hamiltonians (\ref{e1})
was positive and discrete \cite{R2} and it was conjectured that spectral
positivity was a consequence of the invariance of $H$ under the combination of
the space-reflection operator $\cP$ and the time-reversal operator $\cT$. Three
years later, a rigorous proof of spectral positivity was given \cite{R3}. Then,
in 2002 it was demonstrated that the time-evolution operator $U=e^{-iHt}$ for
the Hamiltonian (\ref{e1}) is unitary \cite{R4}. In Ref.~\cite{R4} it was shown
that if the $\cP\cT$ symmetry of a non-Hermitian Hamiltonian is unbroken, then
it is possible to construct a new operator called $\cC$ that commutes with the
Hamiltonian $H$. The Hilbert-space inner product with respect to the $\cC\cP\cT$
adjoint has a positive norm. Also, the operator $U$ is unitary with respect to
the $\cC\cP\cT$ adjoint. Thus, Dirac Hermiticity of the Hamiltonian is not a
necessary requirement of a quantum theory and unbroken $\cP\cT$ symmetry is
sufficient to guarantee that the spectrum of $H$ is real and positive and that
the time evolution is unitary.

In subsequent papers the $\cC$ operators for various quantum-mechanical and
field-theoretic models were calculated \cite{R5,R6,R7,R8,R9}, mostly by using
conventional perturbative methods. It was shown that this operator has a natural
form as the parity operator multiplied by an exponential of a Dirac Hermitian
operator $\cQ$:
\begin{equation}
\cC=e^\cQ\cP,~{\rm where}~\cQ=\cQ^\dag.
\label{e2}
\end{equation}
The operator $\cQ$ vanishes in the unperturbed $\epsilon\to0$ limit when the
Hamiltonian becomes Hermitian and parity invariant. This implies that the $\cC$
operator can be interpreted as the complex extension of the parity operator
$\cP$. It was proved by Mostafazadeh that the $\cQ$ operator can be used to
transform the non-Dirac-Hermitian Hamiltonian $H$ to a spectrally equivalent
Dirac-Hermitian Hamiltonian $h$ \cite{R10}:
\begin{equation}
h=e^{\cQ/2}He^{-\cQ/2}.
\label{e3}
\end{equation}
This similarity transformation was used by Geyer {\it et al.} to convert
Hermitian Hamiltonians to non-Hermitian Hamiltonians \cite{R11}.
 
Originally, Bender and Boettcher introduced the Hamiltonian (\ref{e1}) to
examine the conjecture by Bessis and Zinn-Justin that the spectrum of the
Hamiltonian $H=p^2+ix^3$ might be real. Bender and Boettcher speculated that if
this conjecture were true, then the reality of the spectrum might be due to the
obvious symmetry of this Hamiltonian under combined $\cP$ and $\cT$ reflection.
To study this conjecture Bender and Boettcher considered the Hamiltonian in
(\ref{e1}) because (i) this Hamiltonian is $\cP\cT$ symmetric for all real
$\epsilon$, and (ii) it could then be studied perturbatively for
small $\epsilon$ by using the methods of the $\delta$ expansion, which had been
developed much earlier by Bender {\it et al} \cite{R12}. (One recovers the
Bessis-Zinn-Justin model Hamiltonian by setting $\epsilon=1$.) The discovery
that order-by-order in powers of $\epsilon$ the eigenvalues of $H$ in (\ref{e1})
are {\it all} real led to much subsequent numerical and analytical work on this
model.

Surprisingly, the methods of the $\delta$ expansion, for which the principal
idea is to introduce in the {\it exponent} a small perturbation parameter whose
effect is to quantify how nonlinear a theory is, was not used in further studies
of the Hamiltonian in (\ref{e1}). The objective of this paper is to report a new
perturbative study along these lines in which the $\cC$ operator and the
equivalent Dirac-Hermitian Hamiltonian $h$ are calculated for small $\epsilon$.
We have determined the $\cC$ operator to first order in powers of $\epsilon$,
and using this result we have found the equivalent Hermitian Hamiltonian $h$ to
second order in $\epsilon$. The results can be presented compactly, but they
reveal in dramatic fashion how complicated and nonlocal the isospectral Dirac
Hermitian Hamiltonian $h$ can be.

The construction of the $\cC$ operator in Ref.~\cite{R4} was the key step in
showing that time evolution for the non-Hermitian Hamiltonian (\ref{e1}) is
unitary. However, the difficulty with the construction given in Ref.~\cite{R4}
is that calculating the $\cC$ operator requires as input all of the
coordinate-space eigenvectors of the Hamiltonian. This information is available
in quantum mechanics but it is unwieldy. (In the case of quantum field theory
this information is not available because there is no simple analog of the
coordinate-space Schr\"odinger equation.)

Fortunately, it is possible obtain the $\cC$ operator by solving three simple
simultaneous algebraic equations \cite{R5}:
\begin{equation}
\cC^2=1,
\label{e4}
\end{equation}
\begin{equation}
\left[\cC,\cP\cT\right]=0,
\label{e5}
\end{equation}
\begin{equation}
\left[\cC,H\right]=0.
\label{e6}
\end{equation}
The first two of these equations are kinematic because they are obeyed by the
$\cC$ operator for any $\cP\cT$-symmetric Hamiltonian. If we seek a solution for
$\cC$ in the form (\ref{e2}), we find that these two equations imply that the
operator $\cQ(x,p)$ is an even function of the $x$ operator and an odd function
of the $p$ operator. The third equation (\ref{e6}) is dynamical because it makes
explicit use of the Hamiltonian that defines the theory. This is the equation
that we will solve perturbatively using the methods of the $\delta$ expansion.

In Sec.~\ref{s2} we calculate the $\cQ$ operator to first order in $\epsilon$
for the $\cP\cT$-symmetric Hamiltonian in (\ref{e1}) and in Sec.~\ref{s3} we
calculate the equivalent Hermitian Hamiltonian $h$ to second order.
In Sec.~\ref{s4} we discuss the calculation of eigenvalues.

\section{First-Order Calculation of the $\cC$ Operator}
\label{s2}

We begin our calculation of the $\cC$ operator for $H$ in (\ref{e1}) by
expanding $H$ to second-order in powers of $\epsilon$:
\begin{equation}
H=H_0+\epsilon H_1+\epsilon^2H_2+{\rm O}(\epsilon^3),
\label{e7}
\end{equation}
where $H_0=p^2+x^2$ is the Hamiltonian for the harmonic oscillator, $H_1=x^2\log
(ix)$, and $H_2=\frac{1}{2}[x\log(ix)]^2$. We then recall the representation of
the $\cC$ operator in (\ref{e2}) and expand the $\cQ$ operator as a series in
powers of $\epsilon$:
\begin{equation}
\cQ=\sum_{n=1}^\infty\epsilon^n\cQ_n.
\label{e8}
\end{equation}
This perturbation series begins at $n=1$ because when $\epsilon=0$, the $\cC$
operator reduces to the parity operator $\cP$. We will see that even-$n$ terms
as well as odd-$n$ terms must be included in (\ref{e8}). This is a significant
departure from previous perturbative results for the $\cC$ operator; for the
cubic Hamiltonian $H=p^2+x^2+\epsilon ix^3$ \cite{R6} and the square-well
Hamiltonian $H=p^2+V(x)$, where $V(x)=\epsilon ix/|x|$ ($|x|<1$) and $V(x)=
\infty$ ($|x|>1$) \cite{R7}, there are no even-$n$ terms.

If we then substitute the expansion for $\cC$,
$$\cC=\left[\mathbf{1}+\epsilon\cQ_1+\epsilon^2\cQ_2+\frac{1}{2}\epsilon^2
\cQ_1^2+{\rm O}(\epsilon^3)\right]\cP,$$
and the expansion for $H$ in (\ref{e7}) into the commutator (\ref{e6}) and
collect powers of $\epsilon$, we obtain a sequence of equations for the
coefficients of $\cQ$. After some algebra, we find that the first-order
equation simplifies to
\begin{equation}
\left[\cQ_1,H_0\right]=x^2\left[\log(ix)-\log(-ix)\right]=i\pi x|x|
\label{e9}
\end{equation}
and that the second-order equation becomes
\begin{equation}
\left[\cQ_2,H_0\right]=i\pi x|x|\log|x|-\left[\cQ_1,x^2\log|x|\right].
\label{e10}
\end{equation}

Although we have not yet found an analytical solution to (\ref{e10}), it is
likely from this equation that $\cQ_2$ is nonzero, and as we stated earlier,
this is an unexpected result based on previous perturbative calculations of the
$\cC$ operator. Although (\ref{e9}) is simple looking, it is difficult to solve.
Nevertheless, we have found an exact analytical solution.

The solution of (\ref{e9}) relies heavily on the work of Bender and Dunne
\cite{R13}. We introduce the set of Weyl-ordered operators
\begin{equation}
T_{m,n}\equiv\frac{1}{2^m}\sum_{k=0}^m\binom{m}{k}p^k x^n p^{m-k}\quad(m,\,n=
0,\,1,\,2,\,\ldots).
\label{e11}
\end{equation}
The operator $T_{m,n}$ is a totally-symmetric quantum-mechanical generalization
of the classical product $p^mx^n$. Weyl-ordered operator products rely
implicitly on the Heisenberg algebraic property that $[x,p]=i$. We then define
the generalized Weyl-ordered operator $\widetilde{T}_{m,n}$ as
\begin{equation}
\widetilde{T}_{m,n}\equiv\frac{1}{2^m}\sum_{k=0}^m\binom{m}{k}p^k|x|^np^{m-k}.
\label{e12}
\end{equation}
In terms of this definition of Weyl ordering, we assert that $\cQ_1$ can be
expressed as
\begin{equation}
\cQ_1=\displaystyle\frac{\pi}{2}\sum_{n=0}^\infty\frac{(-1)^n}{(2n-1)(2n+1)}
\widetilde{T}_{2n+1,-2n+1}.
\label{e13}
\end{equation}
It is clear that $\cQ_1$ satisfies the kinematic constraints that it be even in
$x$ and odd in $p$. We do not claim that (\ref{e13}) is the unique solution to
(\ref{e9}).

To show that $\cQ_1$ solves (\ref{e9}), we first demonstrate that the $n=0$ term
in the series commuted with $x^2$ gives $i\pi x|x|$; that is,
\begin{equation}
-\frac{\pi}{2}\left[\widetilde{T}_{1,1},x^2\right]=i\pi x|x|.
\label{e14}
\end{equation} 
We then show that the $k$th term in the series commuted with $p^2$ is exactly
canceled by the $(k+1)$st term commuted with $x^2$.

To verify (\ref{e14}) we note that the $n=0$ term in the series is $-\frac{\pi}
{2}\widetilde{T}_{1,1}=-\frac{\pi}{4}\left(|x|p+p|x|\right)$. We then get
$$\left[-\frac{\pi}{4}\left(|x|p+p|x|\right),x^2\right]=-\frac{\pi}{4}\left(|x|
\left[p,x^2\right]+\left[p,x^2\right]|x|\right)=i\pi x|x|.$$

Next we show that the $k$th term of $\cQ_1$ commuted with $p^2$ gives
\begin{equation}
\frac{\pi}{2}\left[\frac{(-1)^k}{(2k-1)(2k+1)}\widetilde{T}_{2k+1,-2k+1},p^2
\right]=-\frac{i\pi(-1)^k}{2^{2k+2}(2k+1)}\sum_{j=0}^{2k+2}\binom{2k+2}{j}p^j
{\rm sgn}(x)x^{-2k}p^{2k+2-j},
\label{e15}
\end{equation}
where the sign function ${\rm sgn}(x)$ is defined by
$${\rm sgn}(x)=\left\{\begin{array}{ll}
1 & \textrm{if $x>0$},\\
0 & \textrm{if $x=0$}, \\
-1 & \textrm{if $x<0$}.
\end{array}\right.$$
The identity $\frac{d}{dx}|x|={\rm sgn}(x)$ implies that $\left[|x|
,p\right]=i\,{\rm sgn}(x)$. Also, from the identity $0=\frac{d}{dx}\left(|x|^n
|x|^{-n}\right)=n|x|^{-1}{\rm sgn}(x)+|x|^n\frac{d}{dx}|x|^{-n}$ for all integer
$n$, we obtain $\left[|x|^{-2k+1},p\right]=i(1-2k){\rm sgn}(x)x^{-2k}$. It
follows that
\begin{eqnarray}
\left[\widetilde{T}_{2k+1,-2k+1},p\right] &=&
\left[\frac{1}{2^{2k+1}}\sum_{j=0}^{2k+1}\binom{2k+1}{j}p^j|x|^{-2k+1}p^{2k+1-j}
,p\right]\nonumber\\
&=& \frac{1}{2^{2k+1}}\sum_{j=0}^{2k+1}\binom{2k+1}{j}p^j\left[|x|^{-2k+1},p
\right]p^{2k+1-j}\nonumber\\
&=& \frac{i(1-2k)}{2^{2k+1}}\sum_{j=0}^{2k+1}\binom{2k+1}{j}p^j{\rm sgn}(x)x^{-2
k}p^{2k+1-j}.\nonumber
\end{eqnarray}
Using this result, we find that
\begin{eqnarray}
&&\left[\widetilde{T}_{2k+1,-2k+1},p^2\right]=p\left[\widetilde{T}_{2k+1,-2k+1},
p\right]+\left[\widetilde{T}_{2k+1,-2k+1},p\right]p\nonumber\\
&=&\frac{i(1-2k)}{2^{2k+1}}\left[\sum_{r=0}^{2k+1}\binom{2k+1}{r}p^{r+1}
{\rm sgn}(x)x^{-2k}p^{2k+1-r}+\sum_{s=0}^{2k+1}\binom{2k+1}{s}p^{s}
{\rm sgn}(x)x^{-2k}p^{2k+2-s}\right]\nonumber\\
&=&\frac{i(1-2k)}{2^{2k+1}}\left[\sum_{r'=1}^{2k+2}\binom{2k+1}{r'-1}p^{r'}{\rm
sgn}(x)x^{-2k}p^{2k+2-r'}\right.+\nonumber\\
&&\qquad\qquad\qquad\qquad\left.+{\rm sgn}(x)x^{-2n}p^{2n+2}+\sum_{s=1}^{2k+1}
\binom{2k+1}{s}p^s{\rm sgn}(x)x^{-2k}p^{2k+2-s}\right]\nonumber\\
&=&\displaystyle\frac{i(1-2k)}{2^{2k+1}}\Bigg[{\rm sgn}(x)x^{-2n}p^{2n+2}+p^{2n+
2}{\rm sgn}(x)x^{-2n}+\nonumber\\
&&\qquad\qquad\qquad\qquad+\sum_{j=1}^{2k+1}\left\{\binom{2k+1}{j-1}+
\binom{2k+1}{j}\right\}p^j{\rm sgn}(x)x^{-2k}p^{2k+2-j}\Bigg].\nonumber
\end{eqnarray}
We then simplify this expression by using
$$\binom{2k+1}{j-1}+\binom{2k+1}{j}=\binom{2k+2}{j},$$
which establishes the result in (\ref{e15}).

We now calculate the $(k+1)$st term of $\cQ_1$ commuted with $x^2$. We first
note that $\widetilde{T}_{m,n}$ can be rewritten as \cite{R13}
$$\widetilde{T}_{m,n}=\frac{1}{2^n}\sum_{j=0}^n\binom{n}{j}|x|^jp^m|x|^{n-j}.$$
Using the identity
$$\left[p^n,x^2\right]=-in\left(p^{n-1}x+xp^{n-1}\right)
=-2inxp^{n-1}+n(n-1)p^{n-2}=-2inp^{n-1}x-n(n-1)p^{n-2},$$
we obtain
\begin{eqnarray}
\left[\widetilde{T}_{m,n},x^2\right] &=& \frac{1}{2^n}\sum_{j=0}^n\binom{n}{j}
|x|^j\left[p^m,x^2\right]|x|^{n-j}\nonumber\\
&=& \frac{1}{2^n}\sum_{j=0}^n\binom{n}{j}|x|^j\left[-im(xp^{m-1}+
p^{m-1}x)\right]|x|^{n-j}\nonumber\\
&=& \frac{-2im}{2^n}\sum_{j=0}^n\binom{n}{j}|x|^j\frac{1}{2}\left(x
p^{m-1}+p^{m-1}x\right)|x|^{n-j}\nonumber\\
&=& \frac{-2im}{2^{m-1}}\sum_{r=0}^{m-1}\binom{m-1}{r}p^r\epsilon(x
)|x|^{n+1}p^{m-1-r}.\nonumber
\end{eqnarray}
Thus,
$$\left[\widetilde{T}_{2k+3,-2k-1},x^2\right]=\frac{-2i(2k+3)}{2^{2k+2}}
\sum_{r=0}^{2k+2}\binom{2k+2}{r}p^r\epsilon(x)x^{-2k}p^{2k+2-r}.$$
This shows that
$$\frac{\pi}{2}\left[\frac{(-1)^k}{(2k-1)(2k+1)}\widetilde{T}_{2k+1,-2k+1},p^2
\right]=-\frac{i\pi(-1)^k}{2^{2k+2}(2k+1)}\sum_{j=0}^{2k+2}\binom{2k+2}{j}p^j
\epsilon(x)x^{-2k}p^{2k+2-j}.$$
We conclude that the $k$th term of $\cQ_1$ commuted with $p^2$ is exactly
canceled by the $(k+1)$th term commuted with $x^2$:
$$\frac{\pi}{2}\left[\frac{(-1)^{k+1}}{(2k+1)(2k+3)}\widetilde{T}_{2k+3,-2k-1},
x^2\right]=-\frac{\pi}{2}\left[\frac{(-1)^k}{(2k-1)(2k+1)}\widetilde{T}_{2k+1,
-2k+1},p^2\right].$$

We have thus shown that $\cQ_1$ in (\ref{e13}) solves the commutation relation
(\ref{e9}). This argument was quite elaborate, and it is clear why we have not
yet found an analytical solution to the commutation relation (\ref{e10}).
However, having found $\cQ_1$, we can now calculate the equivalent Hermitian
Hamiltonian $h$ to {\it second} order in $\epsilon$, as we show in the next
section.

\section{Second-Order Calculation of the Equivalent Hermitian Hamiltonian}
\label{s3}

In this section we use (\ref{e3}) to calculate the Hermitian
Hamiltonian $h$, which is isospectral to $H$ in (\ref{e1}). Expanding
(\ref{e3}) as a perturbation series in powers of $\epsilon$, we obtain
\begin{eqnarray}
h &=& e^{-\frac{\cQ}{2}}He^{\frac{\cQ}{2}}=\left(1-\frac{\cQ}{2}+\frac{\cQ^2}{8}
+\ldots\right)H\left(1+\frac{\cQ}{2}+\frac{\cQ^2}{8}+\ldots\right)\nonumber\\
&=& H_0+\epsilon\left(\frac{1}{2}\left[H_0,\cQ_1\right]+H_1\right)\nonumber\\
&& \quad+\epsilon^2\left(\frac{1}{2}\left[H_0,\cQ_2\right]+\frac{1}{8}\left\{
H_0,\cQ_1^2\right\}+\frac{1}{2}\left[H_1,\cQ_1\right]-\frac{\cQ_1}{2}H_0\frac{
\cQ_1}{2}+H_2\right)+{\rm O}(\epsilon^3),
\label{e16}
\end{eqnarray}
where curly brackets indicate anticommutation relations. We evaluate the first-
and second-order terms in this equation and show that they can be reduced
to compact forms.

\begin{figure*}[t!]
\vspace{2.8in}
\includegraphics{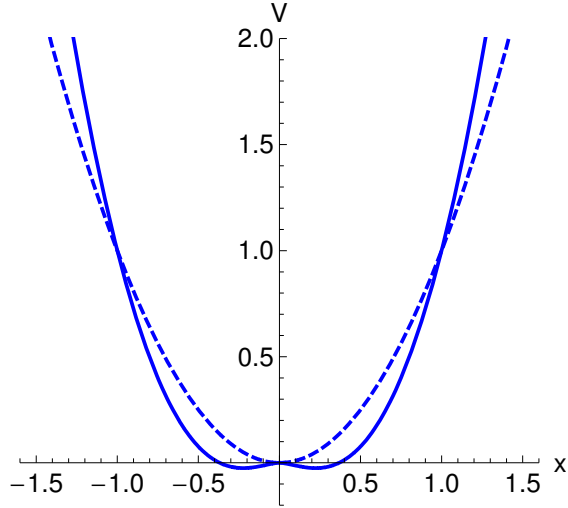}
\caption{Two plots of the potential $V(x)=x^2+\epsilon x^2\log|x|$, one for the
unperturbed case $\epsilon=0$ (dashed line) and the other for $\epsilon=1$
(solid line). Note that when $\epsilon>0$, the potential $x^2+\epsilon x^2\log|x
|$ lies below $x^2$ for $|x|<1$, but for $|x|>1$ it rises faster than the
parabolic potential and thus squeezes the energy levels upward.}
\label{f1}
\end{figure*}

To first order in $\epsilon$, we use (\ref{e9}) to simplify (\ref{e16}) and get
\begin{equation}
h=H_0+\epsilon\left(-\frac{i\pi}{2}x|x|+x^2\log(ix)\right)=
p^2+x^2+\epsilon x^2\log|x|.
\label{e17}
\end{equation}
Thus, to first order in $\epsilon$ the potential for the equivalent Hermitian
Hamiltonian $h$ is a minor correction to the potential $x^2$ for the harmonic
oscillator. As shown in Fig.~\ref{f1}, when $\epsilon$ is positive, the
potential $x^2+\epsilon x^2\log|x|$ lies below $x^2$ for $|x|<1$,
but for $|x|>1$ it rises faster than the parabolic potential and thus
squeezes the energy levels upward.

To second order in $\epsilon$, the equivalent Hermitian Hamiltonian takes the
form
$$h=H_0+\epsilon x^2\log|x|+\epsilon^2 f(x,p)+{\rm O}(\epsilon^3),$$
which can be simplified by using (\ref{e10}): 
\begin{eqnarray}
f(x,p) &=& \frac{1}{2}\left[H_0,\cQ_2\right]+\frac{1}{8}\left\{H_0,\cQ_1^2
\right\}+\frac{1}{2}\left[H_1,\cQ_1\right]-\frac{\cQ_1}{2}H_0\frac{\cQ_1}{2}
+H_2\nonumber\\
&=&
\frac{1}{2}\left(\left[\cQ_1,H_1\right]+\left[\frac{1}{2}\cQ_1^2,H_0\right]-i\pi
x|x|\log|x|-\cQ_1\left[\cQ_1,H_0\right]\right)\nonumber\\
&& \qquad\qquad\qquad\qquad+\frac{1}{8}\left\{H_0,\cQ_1^2\right\}+
\frac{1}{2}\left[H_1,\cQ_1\right]-\frac{1}{4}\cQ_1H_0\cQ_1+H_2\nonumber\\
&=& \frac{1}{4}\left[\cQ_1^2,H_0\right]-\frac{1}{2}\cQ_1\left[\cQ_1,H_0\right]+
\frac{1}{8}\left\{H_0,\cQ_1^2\right\}-\frac{1}{4}\cQ_1H_0\cQ_1\nonumber\\
&& \qquad\qquad\qquad\qquad+H_2-\frac{i\pi}{2}x|x|\log|x|.
\label{e18}
\end{eqnarray}
The significance of this formula is that we do not need $\cQ_2$ to calculate $h$
to order $\epsilon^2$.

The result in (\ref{e18}) may be further simplified by using
$$H_2-\frac{i\pi}{2}x|x|\log|x|=\frac{1}{2}x^2\left[\log(ix)\right]^2-\frac{1}{
2}i\pi x|x|\log|x|=\frac{1}{2}x^2\left(\log|x|\right)^2-\frac{\pi^2}{8}x^2$$
and
\begin{eqnarray}
&& \frac{1}{4}\left[\cQ_1^2,H_0\right]-\frac{1}{2}\cQ_1\left[\cQ_1,H_0\right]+
\frac{1}{8}\left\{H_0,\cQ_1^2\right\}-\frac{1}{4}\cQ_1H_0\cQ_1\nonumber\\
&& \quad=\frac{1}{4}\cQ_1^2H_0-\frac{1}{4}H_0\cQ_1^2-\frac{1}{2}
\cQ_1^2H_0+\frac{1}{2}\cQ_1H_0\cQ_1
+\frac{1}{8}H_0\cQ_1^2+\frac{1}{8}\cQ_1^2H_0-\frac{1}{4}\cQ_1H_0Q_1\nonumber\\
&& \quad=-\frac{1}{8}\cQ_1^2H_0+\frac{1}{4}\cQ_1H_0\cQ_1-\frac{1}{8}H_0Q_1^2
\nonumber\\
&& \quad=\frac{1}{8}\left[\left[\cQ_1,H_0\right],\cQ_1\right].
\nonumber
\end{eqnarray}
Thus, the second-order correction can be written as
\begin{equation}
f(x,p)=\frac{1}{2}x^2\left(\log|x|\right)^2-\frac{\pi^2}{8}x^2+\frac{1}{8}\left[
\left[\cQ_1,H_0\right],\cQ_1\right].
\label{e19}
\end{equation}

We then calculate the double commutator $\left[\left[\cQ_1,H_0\right],\cQ_1
\right]$ as follows:
\begin{eqnarray}
\left[\left[\cQ_1,H_0\right],\cQ_1\right] &=& i\pi\left[x|x|,\cQ_1\right]
\nonumber\\
&=& i\pi\left\{\begin{array}{ll}
\left[x^2,\cQ_1\right] & \textrm{if $x>0$}\\
-\left[x^2,\cQ_1\right] & \textrm{if $x<0$}\end{array}\right.\nonumber\\
&=& i\pi\left\{\begin{array}{ll}
\frac{\pi}{2}\sum_{n=0}^\infty\frac{(-1)^n}{(2n-1)(2n+1)}\left[x^2
,T_{2n+1,-2n+1}\right] & \textrm{if $x>0$}\\
\frac{\pi}{2}\sum_{n=0}^\infty\frac{(-1)^n}{(2n-1)(2n+1)}\left[x^2
,T_{2n+1,-2n+1}\right] & \textrm{if $x<0$}\end{array}\right.\nonumber\\
&=& \frac{i\pi^2}{2}\sum_{n=0}^\infty\frac{(-1)^n}{(2n-1)(2n+1)}
\left[x^2,T_{2n+1,-2n+1}\right].\nonumber
\end{eqnarray}
Recalling that $\left[x^2,T_{2n+1,-2n+1}\right]=2i(2n+1)T_{2n,-2n+2}$
\cite{R13}, we get
$$f(x,p)=\frac{1}{2}x^2(\log|x|)^2-\frac{\pi^2 x^2}{8}-\frac{\pi^2}{8}
\sum_{n=0}^\infty~\frac{(-1)^n}{2n-1}T_{2n,-2n+2}.$$

A major result of Ref.~\cite{R13} is that
$$\sum_{n=0}^\infty~\frac{(-1)^n}{2n+1}T_{2n+1,-2n-1}=\arctan\left(p\frac{1}
{x}\right)-iF(x,p),$$
where
$$F(x,p)=\frac{1}{2H_0}\int_0^\infty
ds\,\frac{e^s}{\cosh(\frac{s}{2H_0})}.$$
We simply take the Dirac-Hermitian conjugate of this equation, which inverts the
order of the operators $p$ and $x$. The definition of
Weyl-ordered operators $T_{m,n}$ guarantees that they are Dirac-Hermitian.
Then, noting that $F(x,p)^\dag=F(x,p)$, we deduce that
$$\sum_{n=0}^\infty~\frac{(-1)^n}{2n+1}T_{2n+1,-2n-1}=\arctan\left(\frac{1}
{x}p\right)+iF(x,p).$$
Hence,
$$\sum_{n=0}^\infty\frac{(-1)^n}{2n+1}T_{2n+1,-2n-1}=\frac{1}{2}\arctan\left(p
\frac{1}{x}\right)+\frac{1}{2}\arctan\left(\frac{1}{x}p\right).$$

Finally, using $\left\{x,T_{m,n}\right\}=2T_{m,n+1}$ and $\left\{p,T_{m,n}
\right\}=2T_{m+1,n}$ \cite{R13}, we obtain
\begin{eqnarray}
\sum_{n=0}^\infty\frac{(-1)^n}{2n-1}T_{2n,-2n+1}
&=&-\sum_{m=-1}^\infty\frac{(-1)^m}{2m+1}T_{2m+2,-2m}\nonumber\\
&=&-x^2-\sum_{m=0}^\infty\frac{(-1)^m}{2m+1}T_{2m+2,-2m}\nonumber\\
&=&-x^2-\frac{1}{4}\left\{x,\left\{p,\sum_{m=0}^\infty
\frac{(-1)^m}{2m+1}T_{2m+1,-2m-1}\right\}\right\}\nonumber\\
&=&-x^2-\frac{1}{8}\left\{x,\left\{p,\arctan\left(p\frac{1}{x}
\right)+\arctan\left(\frac{1}{x}p\right)\right\}\right\},\nonumber
\end{eqnarray}
where the curly brackets indicate anticommutators. Thus, we obtain an explicit
formula for the equivalent Hermitian Hamiltonian $h$ to second order in
$\epsilon$:
\begin{eqnarray}
h &=& H_0+\epsilon x^2\log|x|\nonumber\\
&\quad& +\epsilon^2\left(\frac{1}{2}x^2\left(\log|x|\right)^2+\frac{\pi^2}
{64}\left\{x,\left\{p,\arctan\left(p\frac{1}{x}\right)+\arctan\left(\frac{1}{x}p
\right)\right\}\right\}\right)+{\rm O}(\epsilon^3).
\label{e20}
\end{eqnarray}
Note that $h$ is singular and nonlocal because it contains all positive powers
of $p$ and all negative powers of $x$.

\section{Calculation of energy eigenvalues}
\label{s4}

If we expand (\ref{e1}) as a series in powers of $\epsilon$, we obtain
$$H=p^2+\epsilon x^2\log(ix)+\half\epsilon^2[\log(ix)]^2+\ldots\,,$$
where $\log(ix)=\log(|x|)+\half i\pi\,{\rm sgn}(x)$. We can then use
conventional perturbation techniques to calculate the ground-state energy as a
series in powers of $\epsilon$:
\begin{equation}
E_{\rm ground~state}=1+a\epsilon+b\epsilon^2+{\rm O}\left(\epsilon^3\right),
\label{e21}
\end{equation}
In first-order perturbation theory the coefficient $a$ is the expectation value
of the perturbing potential $\epsilon x^2\log(ix)$ in the unperturbed
harmonic-oscillator ground state, whose wave function is $\exp(-x^2/2)$. We
find that
\begin{equation}
a=\frac{1}{4}\Psi\left(\frac{3}{2}\right)=\frac{1}{2}-\frac{\gamma}{4}
-\frac{1}{2}\log2=0.009\,122\,49\ldots\,.
\label{e22}
\end{equation}
The same result for $a$ is obtained by truncating the equivalent Hermitian
Hamiltonian $h$ in (\ref{e20}) after the first-order term.

We calculate the coefficient $b$ in (\ref{e21}) by using second-order
perturbation theory applied to $H$. The calculation is lengthy, and we do not
discuss it here except to give the result:
\begin{equation}
b=\frac{1}{128}[8+2\pi^2\log2+4(\gamma-2+2\log2)(\gamma+2\log2)+7\zeta(3)]
=0.232\,89\ldots,
\label{e23}
\end{equation}
where $\gamma=0.5772\ldots$ is Euler's constant.

Note that the ground-state energy of the non-Hermitian Hamiltonian $H=p^2+x^2
(ix)^\epsilon$ is slightly higher than the ground-state energy of the harmonic
oscillator. This is consistent with previous numerical calculations in
Ref.~\cite{R2} and agrees with a precise numerical calculation of the
ground-state energy for the non-Hermitian Hamiltonian (\ref{e1}) when $\epsilon$
is small, as is shown in Table I.

\begin{center}
\begin{table}
\label{t1}
\begin{tabular}{c | c | c | c }
$\boldsymbol{\epsilon}$ & {\bf Numerical value} & {\bf First order} & {\bf
Second order}\\
\hline
0.1 & $1.003\,097$ & $1.000\,912$ & $1.003\,241$ \\
0.01 & $1.000\,114\,36$ & $1.000\,091\,22$ & $1.000\,114\,51$ \\
0.001 & $1.000\,009\,355\,38$ & $1.000\,009\,122\,49$ & $1.000\,009\,355\,22$ \\
\end{tabular}
\caption{Comparison of the exact ground-state eigenvalue for the non-Hermitian
Hamiltonian in (\ref{e1}) and the first- and second-order perturbative
calculations of the ground-state eigenvalue as given in (\ref{e21}--\ref{e23}).}
\end{table}
\end{center}

Because the second-order equivalent Dirac-Hermitian Hamiltonian $h$ in
(\ref{e20}) is nonlocal, it is not clear how to use $h$ to calculate the
energy eigenvalues beyond first order. If we try to do so by using standard
Rayleigh-Schr\"odinger methods, we are faced with the problem of calculating the
expectation value of the arctangent functions in (\ref{e20}). This is a singular
and ill-defined calculation that requires the introduction of a regulator. The
singular nature of this calculation can be seen from the identity $\frac{1}{x}p
|0\rangle=i|0\rangle$. This identity gives the formal result ${\rm arctan}\left(
\frac{1}{x}p\right)|0\rangle={\rm arctan}(i)|0\rangle$, which is infinite in the
absence of a regulator. In fact, the singularity is even more severe than this
observation suggests: If we attempt to evaluate the expectation value of the
arctangent functions without a regulator and use the Taylor expansion 
$${\rm arctan}(s)=\sum_{k=0}^\infty (-1)^k s^{2k+1}/(2k+1),$$
we can only recover the value of $b$ in (\ref{e23}) if we can interpret the
sum of the divergent series $0+1+0+1+0+1+0+\ldots$ as $\half\log(2)$.

We conclude that, while the equivalent isospectral Hamiltonian $h$ is formally
Dirac Hermitian, using $h$ to calculate the energies presents serious
difficulties. We have encountered here the same kind of difficulties that were
discovered in Ref.~\cite{R15}, namely, that the Dirac-Hermitian Hamiltonian $h$
suffers from nonlocality and consequently is hard to use in a calculation.
In Ref.~\cite{R15} dimensional regulation was used to
eliminate divergences in Feynman diagrams. We defer to a future paper the search
for a suitable regulator for the present problem.

\acknowledgments{We thank Dr.~D.~Hook for his assistance with numerical
calculations. CMB is grateful to the Theoretical Physics Group at Imperial
College for its hospitality and he thanks the U.S. Department of Energy for
financial support.}

\end{document}